\documentstyle[sprocl]{article}

\def\Journal#1#2#3#4{{#1} {\bf #2}, #3 (#4)}

\def\SS{\em Surf. Sci.}
\def\PRL{\em Phys. Rev. Lett.}
\def\PRB{{\em Phys. Rev.} B}
\def\ZPB{{\em Z. Phys.} B}

\def\vep{\varepsilon}
\def\ra{\rightarrow}
\def\be{\begin{equation}}
\def\ee{\end{equation}}
\def\bea{\begin{eqnarray}}
\def\eea{\end{eqnarray}}

\begin{document}

\title{COULOMB EFFECTS ON THE MAGNETOCONDUCTANCE OF A TWO-DIMENSIONAL 
ELECTRON GAS IN A LATERAL SUPERLATTICE: A SCREENED HARTREE-FOCK
CALCULATION} 

\author {A. MANOLESCU}
\address{
Institutul de Fizica \c{s}i Tehnologia Materialelor, C.P. MG-7
Bucure\c{s}ti-M\u{a}gurele, Rom\^ania}
\author{R. R. GERHARDTS}
\address{
Max-Planck-Institut f\"ur Festk\"orperforschung, Heisenbergstrasse 1,
D-70569 Stuttgart, Federal Republic of Germany}

\maketitle\abstracts{
We calculate the magnetoconductivity tensor of a 2D electron 
gas in a 1D periodic potential and quantising magnetic fields. 
We study the internal structure of the Shubnikov-de Haas
peaks and analyse recent experimental results.  The electron-electron 
interaction is accounted for within a screened Hartree-Fock 
approximation (HFA), which describes both compressible strips 
and an exchange-enhanced spin splitting, unlike the Hartree and 
the standard HFA.}

\section{Introduction}

Recent magnetotransport experiments on GaAs-AlGaAs
heterostructures with a built-in 1D lateral electric modulation, 
in the $x$ direction, have focused on the internal structure of the 
Shubnikov-de Haas peaks of the resistivity $\rho_{xx}$.~\cite{WKP,MGT} 
This structure is determined by the density of 
states (DOS) of the 1D Landau bands and by their exchange-enhanced 
spin splitting.  The interpretation of the experimental
results is however difficult and still unclear.

The electron-electron interaction has  important effects.
The electrostatic and the exchange components act oppositely: 
The Hartree screening reduces the energy dispersion
of the Landau bands, {\em increasing} the DOS, 
while the negative exchange energy of the occupied states
broadens the bands, {\em decreasing} the DOS.~\cite{MG}
Our previous attempt to take into account such effects in a
magnetotransport calculation suggested that the standard
Hartree-Fock approximation (HFA) overestimates the exchange.~\cite{MGT} 
The resulting van Hove singularities (VHS) of the DOS become 
very sharp, and for  modulation periods much 
larger than the magnetic length $l$ the competition between 
the long-range electrostatic screening and  the exchange broadening may 
result in unrealistic short-range oscillations of the charge density. 

In the present paper we overcome such artefacts by using a
screened HFA (SHFA), i.e.~we include the screening influence on the
exchange interaction. We show that the enhanced spin splitting 
can coexist with the picture of compressible/incompressible strips.
Our modulation model is a simple potential $V\!\cos Kx$, uniform 
on the $y$ axis. The material parameters are those for GaAs.

\section{Screened Hartree-Fock Approximation}

Our one-particle Hamiltonian has the form
$H = H^0 + \Sigma^{ee} +\Sigma^{ei}$, in which
$H^0$ describes an electron in the plane $\{{\bf r}=(x,y)\}$,
in a perpendicular magnetic field $B$ and in the periodic potential,
while $\Sigma^{ee}$ and $\Sigma^{ei}$ are the electron-electron
and electron-impurity self-energies.  We use the Hartree-Fock 
expression for $\Sigma^{ee}$, at a finite temperature $T$, 
and we include the polarisation loop in the interaction line 
of the exchange diagram by replacing the Coulomb potential 
$u(q)=2\pi/q$ by $u(q)/\vep(q)$. Here $\vep(q)$  is the 
{\it static} dielectric function of the 2D electron gas,
which we consider quasi-homogeneous, characterised by
$lV/a\ll \hbar\omega_c$, $a=2\pi/K$. 
We calculate $\vep(q)$, for an arbitrary $q$, in the
spirit of the random-phase approximation, 
using the Lindhard formula self-consistently 
with the eigenstates of the Hamiltonian $H$, within a numerical
iterative scheme. The dominant screening corresponds to $ql\ll 1$ 
and is due to the intra-band transitions, which are determined 
by the DOS at the Fermi level, $D_F$. In our SHFA, charge-density 
instabilities of the homogeneous system ($V=0$) are no longer 
possible, in contrast to the standard HFA.

In Fig.1 we show typical results in the limit of isolated wires, 
created by a long-period modulation.  This picture combines the 
specific aspects of both Hartree~\cite{CSG,LG} 
and Hartree-Fock~\cite{MG,DGH} approximations. The former 
leads to strong screening, but only bare spin splitting 
(very small for GaAs), the latter to strong exchange 
enhancement, but poor screening.  In the SHFA the spin
splitting of the compressible edge states is accomplished 
by local fluctuations of the Landau bands with opposite spins,
and of the spin density, but each band is individually pinned 
at the Fermi level.  This result disagrees with the prediction 
of Dempsey et al.,~\cite{DGH} that when the lateral confinement 
decreases, a sharp transition from spin unpolarised to spin polarised 
edge states, with a steep  energy dispersion at the Fermi level, 
should occur. We believe this prediction is related to the 
artefacts of the HFA.  Note that, due to the exchange effects we can 
obtain stable incompressible strips in the bulk of the wires, 
even without impurity broadening.

\section{Conductivities}

We consider the electron-impurity interaction in a phenomenological
self - consistent Born approximation, taking $\Sigma^{ei}$ as
a c-number determined by a characteristic energy parameter
$\Gamma=\gamma\sqrt{B{\rm [Tesla]}}$ [meV].  
We calculate the conductivities using the Kubo formalism adapted
to our system,~\cite{ZG} in which we define the velocity operators
as ${\bf v}=i [H,{\bf r}]/\hbar$.  The only contribution
of the self-energy to the commutator is that of the (nonlocal)
exchange interaction, which is in fact the current vertex correction
required by the Ward identity.~\cite{G}

The relation between the conductivities and the DOS is complicated.
Both the diagonal conductivities $\sigma_{xx}$ and $\sigma_{yy}$
have inter-band-scattering components,  proportional to 
$(\Gamma D_F)^2$ and sensitive to the VHS of the 1D Landau bands, 
when the impurity broadening $\Gamma$ is small. 
Due to the anisotropy  of the system,
only $\sigma_{yy}$ has an intra-band term,  related to the 
dispersion of the Landau band, or, classically, related to the 
drift of the electronic orbits perpendicular to the modulated
electric field.  This  band conductivity is, contrary to
the one due to scattering, approximately proportional to 
$(\Gamma D_F)^{-2}$, thus vanishing near the VHS, but
dominating in $\sigma_{yy}$ for $\Gamma\ra 0$.

For a comparison with the experiment we need to invert the 
conductivities into resistivities, $\rho_{xx,yy}=
\sigma_{yy,xx}/(\sigma_{xx}\sigma_{yy}+\sigma_{xy}^2)$,
and usually $\sigma_{xx}\sigma_{yy}\ll\sigma_{xy}^2$.
For the results presented in Fig.2a we have chosen a 
sufficiently large impurity broadening, so that we have 
a small band conductivity and consequently
$\sigma_{xx}\approx\sigma_{yy}$. In this case the spin 
splitting is not resolvable even without modulation.
For a finite modulation amplitude the Shubnikov-de Haas
oscillations of both longitudinal resistivities may have the 
double-peak structure of the DOS. Such a situation has been 
observed by  Weiss et al.~\cite{WKP} for $\rho_{xx}$, 
in the second Landau band, $n=1$, while here we get it
more clearly in the third band, $n=2$.  At higher magnetic
fields the stronger screening reduces the band width and the
resolution of the VHS is rather poor. The spin-splitting is resolved
in Fig.2b where both modulation amplitude and impurity broadening
are small. For $n=2$ the spin splitted Landau bands still
partly overlap.  The VHS are now observable only in $\rho_{yy}$, 
and are covered by the band conductivity in $\rho_{xx}$. 
Recent measurements have shown a more complicated, double- and
triple-peak profile in $\rho_{xx}$, which may be attributed
to a combined effect of VHS and band conductivity.~\cite{MGT}
But as we have shown, screening effects might be very strong.
The screening can be reduced for a modulation with a shorter period, 
comparable to $l$. The  steeper energy dispersion may allow 
the resolution of the VHS when the band conductivity is 
suppressed by disorder, as suggested in the experiments 
by Sfaxi et al.~\cite{SPL}
However, a comparative measurement of both $\rho_{xx}$ and
$\rho_{yy}$, in high magnetic fields, indicating the scattering
and the band conductivity contributions is, to our knowledge,
not available.

In conclusion, our SHFA describes both screening and exchange-enhanced 
spin splitting of the Landau bands, interpolating in the expected 
manner between the contradictory results of the Hartree and the 
standard Hartree-Fock approximations.  The theoretical 
resistivities are in qualitative agreement with the available 
experimental data. Calculational details will be published 
elsewhere.

\section*{Acknowledgments}

We are grateful to Gabriele Ernst, Behnam Farid, 
Marc Tornow and Dieter Weiss for stimulating discussions.

\newpage

\input{epsf}

\begin{figure}
\epsffile{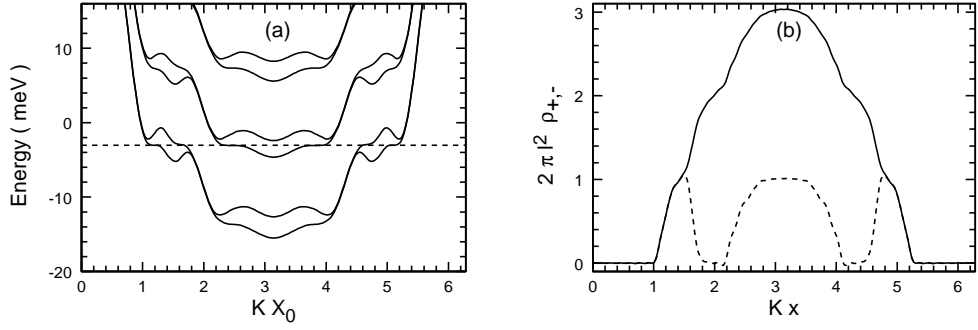}
\caption{(a) \, Landau bands for isolated wires produced by a
modulation with $V=300$~meV and $a=1000$ nm; the dashed 
line shows the Fermi level and $X_0$ is the center coordinate.
(b) \, Particle and spin densities, $\rho_{+,-}$, full
and dashed lines. $B=6$~T and $T=0.5$~K.  
}
\end{figure}

\begin{figure}
\epsffile{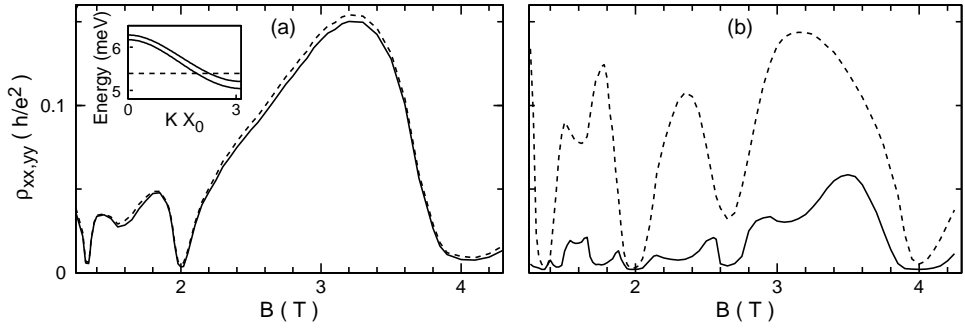}
\caption{The resistivities $\rho_{xx}$, dashed lines, and
$\rho_{yy}$, full lines, for a modulation period $a=300$ nm.
(a) \, $V=12$ meV, $\gamma=0.3$, $T=4.2$ K.  Inset: Landau bands
with $n=1$ for $B=2.9$ T. (b) \, $V=6$ meV, $\gamma=0.05$,
$T=1$ K.
}
\end{figure}

\end{document}